\documentclass[aps,pra,twocolumn,superscriptaddress]{revtex4-1}

\usepackage{soul}
\usepackage{relsize}
\usepackage{lmodern}
\usepackage{slantsc}
\usepackage{graphicx}
\usepackage{amsmath}
\usepackage{amssymb}
\usepackage{amsthm}
\usepackage{amsbsy}
\usepackage{mathrsfs}
\usepackage{varioref}
\usepackage{dsfont}
\usepackage{bm}
\usepackage{color}
\usepackage[usenames,dvipsnames]{xcolor}
\definecolor{myblue}{rgb}{0.2,0.2,0.8}
\definecolor{myblack}{rgb}{0,0,0}
\definecolor{myurl}{rgb}{0.1,0.1,0.4}
\usepackage[colorlinks=true,citecolor=myblue,linkcolor=myblack,urlcolor=myurl]{hyperref}
\usepackage{cleveref}
\usepackage[font = footnotesize, labelfont = bf, justification = RaggedRight]{caption}
\usepackage{subfigure}
\usepackage{booktabs} 
\usepackage{array} 
\usepackage{paralist} 
\usepackage{verbatim} 
\edef\restoreparindent{\parindent=\the\parindent\relax}
\usepackage[parfill]{parskip}
\restoreparindent
\usepackage{natbib}

\setul{}{20 pt}

\newcommand{\ts}{\textsuperscript}
\newcommand{\<}{\langle}
\renewcommand{\>}{\rangle}

\newcommand{\e}{{\mathcal{E}}}

\newcommand{\qq}{{\mathcal{Q}}}

\newcommand{\dd}{{\mathscr{D}}}
\newcommand{\ii}{{\mathcal{I}}}

\newcommand{\co}{\mathds{C}}
\newcommand{\re}{\mathds{R}}

\newcommand{\h}{{\mathcal{H}}}

\newcommand{\one}{\mathds{1}}
\newcommand{\zero}{\mathds{O}}

\newcommand{\tr}{\mathrm{tr}}

\newcommand{\ddh}{\dd^{H}}
\newcommand{\ddpsi}{\dd^{\{\psi\}}}

\newcommand{\ddrho}{\dd^{\rho}}

\newcommand{\eq}[1]{\eqref{#1}}
\newcommand{\fig}[1]{Fig.~\ref{#1}}

\newcommand{\ket}[1]{|{#1}\rangle}

\newcommand{\pr}[1]{P[{#1}]}

\newcommand{\avg}[1]{\langle {#1} \rangle} 
\newcommand{\var}[1]{\mathrm{Var}\left({#1} \right)}

\newcommand{\gmc}{\Gamma}
\newcommand{\qc}{\mathscr{Q}}

\begin{document}

\title{Fluctuating quantum heat}

\author{M. Hamed Mohammady}
\email{m.hamed.mohammady@savba.sk }
\affiliation{RCQI, Institute of Physics, Slovak Academy of Sciences, D\'ubravsk\'a cesta 9, Bratislava 84511, Slovakia}


\begin{abstract}
The increase in average energy of a quantum system undergoing projective energy measurements is referred to as ``quantum heat'', which is always zero. In the framework of quantum stochastic thermodynamics, this is constructed as the average over the fluctuating quantum heat (FQH), defined as the  increase in expected value of the Hamiltonian along two-point eigenstate trajectories. However, such a definition has two drawbacks: (i) if the initial state does not commute with the Hamiltonian and has degeneracies, the higher moments of the FQH will not be uniquely defined, and therefore it is arguable whether such a quantity is physically meaningful; (ii) the definition is operationally demanding as it requires full knowledge of the initial state. In the present manuscript we show that  the FQH is an instance of conditional increase in energy given sequential measurements, the first of which is with respect to the eigen-decomposition of the initial state. By coarse-graining this initial measurement, first by only distinguishing between degenerate subspaces of the state, and finally by not distinguishing between any subspace  at all, we provide two alternative definitions for the FQH, which we call the partially coarse-grained FQH and fully coarse-grained FQH, respectively. The partially coarse-grained FQH resolves issue (i), whereas the fully coarse-grained FQH resolves both (i) and (ii). 
\end{abstract}

\maketitle

\section{Introduction}
When a classical system is brought to thermal equilibrium, by an energy conserving interaction with a thermal reservoir, the increase in  internal energy of the system is the macroscopic heat, and is equal to the  decrease in  internal energy of the reservoir. In the framework of stochastic thermodynamics, the macroscopic heat is the first moment of an underlying fluctuating heat; the evolution of the system's macro-state from an initial to a thermal configuration is decomposed into a probability distribution over trajectories along micro-states, the increase in internal energy of which is the fluctuating, or stochastic, heat \cite{Seifert2008, Sekimoto2010, Seifert2012}. The concept of fluctuating  heat can unambiguously be carried over to quantum systems undergoing thermalisation. Here, the framework of quantum stochastic thermodynamics provides an analogue to classical trajectories as a two-point sequence of energy measurement outcomes $(m,n)$, where   $m$ and $n$ label the outcomes of the initial and final energy measurements, respectively \cite{Campisi2011, Horowitz2013a, Elouard2017b, Manzano2017,  Elouard2018a}.  

However, as argued recently, quantum systems also admit a different kind of heat, referred to as ``quantum heat'', which arise due to projective measurements  \cite{Elouard2017}, and can be used to power quantum thermal machines without the need of a  thermal reservoir \cite{Hayashi2017,Elouard2017a,Mohammady2017,Elouard2018b,Buffoni2018,Solfanelli2019}.   A special instance of such quantum heat is when the observable measured is the Hamiltonian itself. This quantum heat can also be shown to  arise when a system  reaches thermal equilibrium  with a thermal reservoir, and can be understood as the difference between the classical heat and the change in internal energy \cite{Santos2019,Mohammady2019d}. Of course, when a system is projectively measured with respect to its Hamiltonian, the change in average internal energy -- the ``average'' quantum heat -- vanishes for all initial states $\rho$. However, the question of what the underlying fluctuating quantum heat (FQH) should be is not so clear. 

If we extend the notion of fluctuating internal energy from a definite quantity, i.e., an eigenvalue of the Hamiltonian, to the expected value of the Hamiltonian on \emph{any} state of the system, then the standard framework of two-point eigenstate trajectories would give the FQH along the eigenstate trajectory $\gamma := (m,n) \equiv \ket{\psi_m} \mapsto \ket{\varphi_n}$ as $\qq(\gamma) := \< \varphi_n| H |\varphi_n\> - \<\psi_m| H | \psi_m\>$. Here,  $\{\ket{\psi_m}\}$ and $\{\ket{\varphi_n}\}$ are the eigen-decompositions of the   initial and final states, respectively, and so the trajectory $\gamma$ is operationally defined as the sequence of measurement outcomes with respect to such decompositions.   Of course, since the final state has been decohered with respect to the Hamiltonian, then  $\ket{\varphi_n}$ has a definite energy, i.e., $\<\varphi_n| H |\varphi_n\>$ is an energy eigenvalue. However, if the initial state is energy-coherent, i.e., it does not commute with the Hamiltonian, then   $\ket{\psi_m}$ will be energy-indefinite, i.e., $\<\psi_m| H |\psi_m\>$ will not be an energy eigenvalue. The consequence of energy-coherence of the initial state on the FQH, as defined by eigenstate trajectories, is that the higher moments $\avg{\qq^k}$, $k >1$,  will not be uniquely defined if the initial state is degenerate, thus admitting infinitely many eigenstate decompositions, and so infinitely many sets of eigenstate trajectories $ \{\gamma\}$. This clearly raises questions as to whether such a quantity is physically meaningful.  Moreover, such a definition will be operationally demanding; while the fluctuating classical heat requires no knowledge of the system's state, and is given purely by the sequence of energy measurement outcomes, the FQH requires detailed knowledge of the initial state, so that measurements with respect to the eigen-decomposition $\{\ket{\psi_m}\}$ can be performed. 

In this manuscript we shall investigate the ambiguity, and operational meaning, of the FQH in the conceptually simple case where the system, initially in state $\rho$, is projectively measured with respect to its Hamiltonian $H$. We shall first show that  the definition for the FQH, as given by  the eigenstate trajectory approach, is a special instance of the change in internal energy, conditional on sequential measurement outcomes, that was introduced in Ref.  \cite{Mohammady2019c}. The first measurement in this sequence is with respect to the eigen-decomposition of the initial state $\rho$.  By partially ``coarse-graining'' this initial measurement, so as to only distinguish between the degenerate subspaces of $\rho$,   we may arrive at  an alternative definition of the FQH, referred to as the partially coarse-grained FQH $\qc$.  While such a definition is still operationally demanding, it will be identical to the ordinary FQH when the state is non-degenerate, but its higher moments $\avg{\qc^k}$ will always be unique. As such,  it can be seen as a more physically meaningful generalisation of the FQH.  Finally,  by not performing any initial measurement on the system at all,  so as to not distinguish between any subspace of $\rho$, we arrive at the fully coarse-grained FQH   $\mathds{Q}$, which is no longer operationally demanding and only relies on the statistics of a single energy measurement.  While these alternative definitions give different moments of quantum heat in general, they do have three universal relationships: (i)  all three definitions agree in the first moment for all states, i.e., $\avg{\qq} = \avg{\qc} = \avg{\mathds{Q}} = 0$ for all $\rho$; (ii) all three definitions agree in all moments if the initial state is energy incoherent, i.e.,  $\avg{\qq^k} = \avg{\qc^k} = \avg{\mathds{Q}^k} = 0$ for all $k$ if $\rho$ commutes with $H$; and (iii) the variance of all three definitions, which is equal to the second moment, obeys the inequality relationship $\var{\mathds{Q}} \leqslant \var{\qc} \leqslant \var{\qq}$ for all $\rho$.

\section{ Setup}

We consider a system with a finite-dimensional  Hilbert space $\h \simeq \co^d$, and denote by $\zero$ and $\one$ the null and identity operators on $\h$, and by $\pr{\psi} \equiv |\psi\>\<\psi|$ the projection on vector $\ket{\psi} \in \h$. For any self-adjoint operator $A = \sum_i \lambda_i P_i$, with $\lambda_i$ the eigenvalues and $\{P_i\}$ the spectral projections such that $P_i P_j = \delta_{i,j} P_i$ and $\sum_i P_i = \one$, we define the projection map $\dd^{A}(\cdot) := \sum_i P_i (\cdot) P_i$. The projection map has the following useful properties: (i) $\dd^{A}(B) = B$ if and only if $[A, B]=\zero$; (ii) for any operator $B$, $[A, \dd^{A}(B)]=\zero$;  and (iii) $\tr[\dd^{A}(B)C] = \tr[B \dd^{A} (C)] = \tr[\dd^{A}(B) \dd^{A}(C)]$ for any operators $B, C$. In the special case where $A = \sum_i \lambda_i \pr{\psi_i}$ is non-degenerate, with $\{\psi_i\}$ an orthonormal basis of $\h$, we shall use the short-hand notation $\ddpsi \equiv \dd^{A}$.

Measurements of the system will be described by ``instruments'' $\ii := \{\ii_x\}$, where $x$ are the measurement outcomes, and $\ii_x$ are completely positive, trace-non-increasing maps so that the probability of observing outcome $x$, given an initial state $\rho$, is $\tr[\ii_x(\rho)]$, while the conditional state of the system, given that outcome $x$ has been observed, is $\rho(x) := \ii_x(\rho)/\tr[\ii_x(\rho)]$ \cite{Heinosaari2011, Busch2016a}.  Following Ref. \cite{Mohammady2019c}, we define the increase in expected value of Hamiltonian $H$,  on a system that is initially prepared in state $\rho$ and measured by the instrument $\ii := \{\ii_x\}$, conditional on observing outcome $x$, as
\begin{align}\label{eq:cond-energy-change}
\Delta \e(x) & := \frac{\tr[H \ii_x(\rho)]}{\tr[\ii_x(\rho)]} - \frac{\frac{1}{2}\tr[\ii_x(H \rho + \rho H)]}{\tr[\ii_x(\rho)]}
\end{align}
if $\tr[\ii_x(\rho)] >0$, and as $\Delta \e(x) := 0$ if $\tr[\ii_x(\rho)]=0$. Here, the first term is the expected value of $H$ given the normalised post-measurement state $\rho(x)$.  The second term, on the other hand, is the real component of the weak value of $H$, given the initial state $\rho$, and post-selected by outcome $x$ of the instrument $\ii$ \cite{Aharonov1988,Haapasalo2011}.  

Let the system be  initially prepared in the possibly degenerate state 
\begin{align}\label{eq:initial-state}
\rho = \sum_{m} q_m \rho_m \equiv \sum_{m, \mu} \frac{q_m}{d_m} \pr{\psi_{m,\mu}}.
\end{align}
 Here, $q_m \geqslant 0$, $\sum_m q_m =1$, while $\rho_m :=P_m/d_m$.  $P_m$ are the spectral projections of $\rho$, whose eigenstates we denote as   $\ket{\psi_{m ,\mu}}$, where $\mu = 1, \dots, d_m$  characterises the degeneracy. We subsequently projectively measure the system with respect to its Hamiltonian $H = \sum_n \epsilon_n \Pi_n$, with $\epsilon_n$ the energy eigenvalues and $\Pi_n$ the spectral projections. The instrument for such a measurement is thus $\ii_n(\cdot) := \Pi_n (\cdot) \Pi_n$, and an unselective measurement of the Hamiltonian prepares the system in the ``decohered'' state 
 \begin{align}\label{eq:final-state}
 \ddh(\rho) := \sum_n \Pi_n \rho \Pi_n =  \sum_{n,\nu} q_{n,\nu}' \pr{\varphi_{n,\nu}}.
 \end{align}
Since $\ddh(\rho)$ commutes with $H$, then the two operators share an orthonormal basis, and so we may always write $\Pi_n = \sum_\nu \pr{\varphi_{n,\nu}}$.  The average quantum heat, defined as the increase in average energy of the system, is thus
\begin{align}\label{eq:avg-heat}
Q := \tr[H (\ddh(\rho) - \rho)] \equiv \tr[(\ddh(H) - H) \rho] = 0.
\end{align}

\section{Alternative definitions of the fluctuating quantum heat}
\subsection{Eigenstate trajectories}

If we are provided with full knowledge of the initial state $\rho$ given by  \eq{eq:initial-state}, which in turn provides us with full knowledge of the decohered state $\ddh(\rho)$ given by \eq{eq:final-state},  we may projectively measure the system, prior and posterior to the measurement of the Hamiltonian, with respect to the eigen-decompositions $\{\ket{\psi_{m,\mu}}\}$ and $\{\ket{\varphi_{n,\nu}}\}$, respectively. Such a sequential measurement is performed by the instrument
\begin{align}\label{eq:eigenstate-instrument}
\ii_\gamma(\cdot) &:= \pr{\varphi_{n,\nu}}\Pi_n\pr{\psi_{m,\mu}}(\cdot)\pr{\psi_{m,\mu}} \Pi_n \pr{\varphi_{n,\nu}}, \nonumber \\
& = \pr{\varphi_{n,\nu}}\pr{\psi_{m,\mu}}(\cdot)\pr{\psi_{m,\mu}} \pr{\varphi_{n,\nu}},
\end{align}
with the sequential outcomes $\gamma := ((m, \mu) , ( n, \nu ))$ defining the eigenstate trajectory $\ket{\psi_{m,\mu}} \mapsto \ket{\varphi_{n,\nu}}$. In the second line, we have used the fact that $\ket{\varphi_{n,\nu}}$ are eigenstates of $\Pi_n$ to infer that $\Pi_n \pr{\varphi_{n,\nu}} =  \pr{\varphi_{n,\nu}} \Pi_n = \pr{\varphi_{n,\nu}} $. Therefore, the probability of observing the trajectory $\gamma$ is given by \eq{eq:initial-state}  and \eq{eq:eigenstate-instrument} to be
\begin{align}\label{eq:prob-traj}
p(\gamma) &:= \tr[\ii_\gamma(\rho)], \nonumber \\
& = \tr[\pr{\varphi_{n,\nu}}\pr{\psi_{m,\mu}}\rho \pr{\psi_{m,\mu}} \pr{\varphi_{n,\nu}}], \nonumber \\
& =  \frac{q_m}{d_m}\tr[\pr{\varphi_{n,\nu}}\pr{\psi_{m,\mu}}], \nonumber \\
&  =  \frac{q_m}{d_m} |\< \varphi_{n,\nu}|\psi_{m,\mu} \>|^2.
\end{align}
The FQH, on the other hand, can be obtained given \eq{eq:cond-energy-change} and \eq{eq:eigenstate-instrument}, which reads as the difference in internal energy between the initial and final eigenstates:
\begin{align}\label{eq:eigenstate-quantum-heat}
\qq(\gamma) &= \tr[ H \pr{\varphi_{n,\nu}}] - \tr[H\pr{\psi_{m,\mu}}], \nonumber \\
&\equiv \epsilon_n - \tr[H\pr{\psi_{m,\mu}}].
\end{align}
To see how we obtain the first line, note that the normalised conditional state after the measurement is just $\rho(\gamma) := \ii_\gamma(\rho)/\tr[\ii_\gamma(\rho)] = \pr{\varphi_{n,\nu}}$, while since the first measurement projects onto a single eigenstate of $\rho$, we have  $\tr[\ii_\gamma(H \rho)] = \tr[\ii_\gamma( \rho H)] = \tr[\ii_\gamma(\rho)] \tr[H \pr{\psi_{m,\mu}}]$. In the second line we simply use the fact that $\ket{\varphi_{n,\nu}}$ is an eigenstate of $\Pi_n$, and so has the definite energy $\epsilon_n$.

Note that for any $a,b \in \re$, 
\begin{align}\label{eq:binomial}
(a-b)^k = \sum_{i=0}^k \binom{k}{i}(-1)^i a^{k-i} b^i,
\end{align} 
where $\binom{k}{i} := k!/(i ! (k-i)!)$ are the binomial coefficients, and that for any $\ket{\psi}$, $\tr[H \pr{\psi}]^k = \<\psi| (\pr{\psi} H \pr{\psi})^k |\psi\>$. Therefore, defining $\ddpsi(\cdot) := \sum_{m,\mu} \pr{\psi_{m,\mu}}( \cdot)\pr{\psi_{m,\mu}} $, then  by Eqs. \eq{eq:prob-traj}, \eq{eq:eigenstate-quantum-heat}, and \eq{eq:binomial} we may compute the $k\ts{th}$ moment  of the FQH as
\begin{align}\label{eq:moment-heat}
\avg{\qq^k} &:=\sum_\gamma p(\gamma) \qq^k(\gamma), \nonumber \\
& = \sum_{\gamma} \frac{q_m}{d_m} \sum_{i=0}^k \binom{k}{i}(-1)^i   \<\varphi_{n,\nu}|\psi_{m,\mu} \> \< \psi_{m,\mu}| \varphi_{n,\nu} \> \nonumber \\
& \qquad   \times  \epsilon_n^{k-i}   \<\psi_{m,\mu}|(\pr{\psi_{m,\mu}} H \pr{\psi_{m,\mu}} )^i|\psi_{m,\mu}\>  \nonumber  \\
& = \sum_{\gamma}  \sum_{i=0}^k \binom{k}{i}(-1)^i  \nonumber \\
&   \qquad \tr[ H^{k-i} (\pr{\psi_{m,\mu}} H \pr{\psi_{m,\mu}} )^i \rho ] \nonumber  \\
& = \sum_{i=0}^k \binom{k}{i}(-1)^i \tr[H^{k-i} \ddpsi(H)^{i} \rho], \nonumber \\
&= \tr[(H - \ddpsi(H) )^k \rho].
\end{align}
 It is simple to confirm that the first moment given by \eq{eq:moment-heat} agrees with \eq{eq:avg-heat}, since when $k=1$ we have
\begin{align}\label{eq:avg-FQH}
\avg{\qq} \equiv \avg{\qq^1} &= \tr[(H - \ddpsi(H) ) \rho], \nonumber \\
& =\tr[H (\rho - \ddpsi(\rho)) ]  = \tr[H (\rho - \rho)] = 0.
\end{align}
Moreover, if the initial state commutes with the Hamiltonian, then $\ddpsi(H) = H$, and so all moments of the FQH vanish. However, if the initial state $\rho$ does not commute with the Hamiltonian, and is degenerate, then there will be infinitely many possible eigen-decompositions $\{\ket{\psi_{m,\mu}}\}$, resulting in different operators $(H - \ddpsi(H))^k$, and so the higher moments $\avg{\qq^k}$ will not be uniquely defined. 

In particular, since $\avg{\qq} = 0$, the variance in the FQH $\var{\qq} := \avg{\qq^2} - \avg{\qq}^2 $ is just the second moment, and reads
\begin{align}\label{eq:variance-eigenstate}
\var{\qq} & =  \tr[(H - \ddpsi(H) )^2 \rho], \nonumber \\
& = \tr[(H^2 -  \ddpsi(H)^2) \rho], \nonumber \\
& = \tr[H^2 \rho] - \sum_{m,\mu} \frac{q_m}{d_m} \tr[H\pr{\psi_{m,\mu}} H \pr{\psi_{m,\mu}}], \nonumber \\
& = \sum_{m,\mu} \frac{q_m}{d_m} \mathscr{I}_{\pr{\psi_{m,\mu}}}(H).
\end{align}
In the second line we have used the fact that $\ddpsi$ is a projection map on the eigenbasis of $\rho$ to infer  $\tr[H \ddpsi(H) \rho] = \tr[\ddpsi(H) H \rho]= \tr[\ddpsi(H)^2 \rho] $. In the final line we introduce the  Wigner-Yanase-Dyson skew information of $\rho$ with reference to $H$,
 $\mathscr{I}_\rho(H):= \tr[H^2 \rho] - \tr[H \rho^{1/2} H \rho^{1/2}]$,    which: is non-negative and vanishes if $\rho$ commutes with $H$;  is convex in $\rho$;   and $\mathscr{I}_\rho(H) = V_\rho(H)$ whenever $\rho$ is a pure state, where $V_\rho(H) := \tr[H^2 \rho] - \tr[H \rho]^2$ is the variance of $H$ in $\rho$  \cite{Wigner1963, Lieb1973, Takagi2018}. This was the same expression for the variance in FQH obtained in  Ref.\cite{Mohammady2019d}. As shown in \fig{fig:bound}, for a simple model where the system is composed of two qubits and initially prepared in a degenerate state that is energy-coherent, the variance in the FQH will vary depending on the choice of eigen-decomposition $\{\ket{\psi_{m,\mu}}\}$.

\subsection{Partially coarse-grained trajectories}

Already from \eq{eq:eigenstate-quantum-heat} we can see that distinguishing between the degeneracy parameter $\nu$ of energy eigenstates $\ket{\varphi_{n,\nu}}$ is unnecessary, as they all have the same energy $\epsilon_n$. However, the degeneracy parameter $\mu$ of $\rho$ eigenstates $\ket{\psi_{m,\mu}}$ leads to a difference in the FQH depending on what eigen-decomposition we have chosen, and is the cause of ambiguity and non-uniqueness of the higher moments. Consequently, let us  now alter our sequential measurements to be given by the instrument
\begin{align}\label{eq:coarse-grained-instrument}
\ii_\gmc(\cdot) := \Pi_n P_m (\cdot) P_m \Pi_n ,
\end{align}
with the outcomes $\gmc := (m, n )$ defining the   ``partially coarse-grained'' trajectories $\rho_m \mapsto  \Pi_n \rho_m \Pi_n/ \tr[\Pi_n \rho_m]$. Such trajectories  only distinguish between the degenerate subspaces of $\rho$, and the degenerate energy subspaces of $H$. We say that these are partially coarse-grained because they still require measurement with respect to the degenerate subspaces of $\rho$ and, as such, still require that we have full knowledge of the initial state.   Therefore, given \eq{eq:initial-state}, and \eq{eq:coarse-grained-instrument}, the partially coarse-grained trajectories have the probability 
\begin{align}\label{eq:coarse-grained-probability}
p(\gmc) &=  \tr[\Pi_n P_m \rho P_m \Pi_n ] \nonumber \\
&= \frac{q_m}{d_m}\tr[\Pi_n P_m] \equiv q_m \tr[\Pi_n \rho_m ],\nonumber \\
\end{align}
where we note that $p(\Gamma) = \sum_{\mu, \nu} p(\gamma)$.

Defining the partially coarse-grained FQH by \eq{eq:cond-energy-change}, with the instrument defined in \eq{eq:coarse-grained-instrument}, gives us 
\begin{align}\label{eqn:general-Q-heat}
\qc(\gmc)& = \epsilon_n -  \frac{\tr[\Pi_n P_m H P_m ]}{\tr[\Pi_n P_m]}. 
\end{align}
Since the normalised conditional state after the sequential measurement is $\rho(\gmc) := \Pi_n \rho_m \Pi_n/ \tr[\Pi_n \rho_m]$, which has support only on  a single degenerate energy subspace of $H$ projected onto by $\Pi_n$, the first term is simply $\epsilon_n$, as is the case in   \eq{eq:eigenstate-quantum-heat}.  However, the second term is now somewhat more complex, and to see how we obtain it, note that since the first measurement in the sequence is given by the projection operator  $P_m$, we have $\tr[\ii_\Gamma (H \rho)] = \tr[\ii_\Gamma(\rho H)]= (q_m/d_m)\tr[\Pi_n P_m H P_m ]$, while $\tr[\ii_\Gamma(\rho)] = (q_m/d_m)\tr[\Pi_n P_m ]$.  

Formally, by use of Eqs. \eq{eq:binomial}, \eq{eq:coarse-grained-probability}, and \eq{eqn:general-Q-heat},  we may express each moment of $ \qc$ as 
\begin{align}\label{eq:general-heat-moment}
\avg{\qc^k} &:=\sum_\gmc p(\gmc) \qc^k(\gmc), \nonumber \\
&= \sum_{m,n} \frac{q_m}{d_m} \tr[\Pi_n P_m]  \nonumber \\
& \qquad  \times \sum_{i=0}^k \binom{k}{i}(-1)^i \frac{\epsilon_n^{k-i} \tr[\Pi_n P_m H P_m ]^i}{\tr[\Pi_n P_m]^{i}}.
\end{align}
This gives the first moment as 
\begin{align}
\avg{\qc} \equiv \avg{\qc^1} &= \sum_{n} \epsilon_n \tr[\Pi_n \rho] - \tr[H \rho], \nonumber \\
& =\sum_{n} \tr[H \rho] - \tr[H \rho] = 0,
\end{align}
in agreement with both \eq{eq:avg-heat} and \eq{eq:avg-FQH}.  Moreover,  if $\rho$ commutes with $H$, then we have $\tr[\Pi_n P_m H P_m ] = \epsilon_n \tr[\Pi_n P_m ]$, and so \eq{eq:general-heat-moment} can be shown to reduce to 
\begin{align}
\avg{\qc^k} &= \sum_{m,n}  \frac{q_m}{d_m}\tr[\Pi_n P_m]\sum_{i=0}^k \binom{k}{i}(-1)^i  \epsilon_{n}^{k-1} \epsilon_n^i, \nonumber \\
& = \sum_n \tr[\Pi_n \rho] \sum_{i=0}^k \binom{k}{i}(-1)^i   \epsilon_{n}^{k-1} \epsilon_n^i , \nonumber \\
&= \sum_n \tr[\Pi_n \rho](\epsilon_n - \epsilon_n)^k = 0,
\end{align} 
and so we may conclude that $\avg{\qq^k} = \avg{\qc^k}=0$ for all $k$ if $\rho$ commutes with $H$.

In particular, defining the projection map $\ddrho(\cdot) := \sum_m P_m (\cdot) P_m$, the variance $\var{\qc} := \avg{\qc^2} - \avg{\qc}^2 $ reads as 
\begin{align}\label{eq:variance-coarse-grained}
\var{\qc} & = \sum_{m,n}\frac{q_m}{d_m}\bigg( \epsilon_n^2 \tr[\Pi_n P_m] -2 \epsilon_n \tr[\Pi_n P_m H P_m ] \bigg)\nonumber \\
& \qquad + \sum_{m,n}  \frac{q_m}{d_m}\frac{\tr[\Pi_n P_m H P_m ]^2}{\tr[\Pi_n  P_m]} , \nonumber \\
& = \tr[H^2 \rho] - 2 \tr[H \ddrho(H) \rho] \nonumber \\
& \qquad + \sum_{m,n} \frac{q_m}{d_m} \frac{\tr[\Pi_n  P_m H P_m ]^2}{\tr[\Pi_n  P_m]}.
\end{align}
By defining $A := P_m \Pi_n$ and $B := P_m H P_m \Pi_n$, so that the Cauchy-Schwarz inequality $\tr[A^* B]^2 \leqslant \tr[A^* A] \tr[B^* B]$ gives us 
\begin{align}\label{eq:Cauchy-Schwarz-inequality}
\frac{\tr[\Pi_n P_m H P_m ]^2}{ \tr[\Pi_n P_m]} \leqslant \tr[\Pi_n (P_m H P_m)^2],
\end{align} 
and recalling that $\tr[H \ddrho(H) \rho] = \tr[ \ddrho(H) H \rho] = \tr[ \ddrho(H)^2 \rho]$, we therefore obtain the following inequality for the variance in partially coarse-grained FQH:
\begin{align}\label{eq:coarse-grained-variance-inequality}
\var{\qc} &\leqslant \tr[(H - \ddrho(H))^2 \rho], \nonumber \\
& = \tr[(H^2 - \ddrho(H)^2) \rho], \nonumber \\
& = \tr[H^2 \rho] - \sum_m \frac{q_m}{d_m}\tr[H P_m H P_m], \nonumber \\
& =  \tr[H^2 \rho] - \sum_m q_m \tr[H \rho_m^{1/2} H \rho_m^{1/2}], \nonumber \\
& = \sum_m q_m \mathscr{I}_{\rho_m}(H),
\end{align}
where we  recall that $\rho_m = \frac{1}{d_m} P_m $.  Note that  the equality condition of \eq{eq:coarse-grained-variance-inequality} will be satisfied if $P_m \Pi_n$ and $P_m H P_m \Pi_n$ are linearly dependent, which is satisfied either if $P_m$ are all rank-1 projection operators (i.e. $\rho$ is non-degenerate) or if all $P_m$ commute with  $H$ (i.e. $\rho$ commutes with $H$).  Comparing \eq{eq:coarse-grained-variance-inequality} with \eq{eq:variance-eigenstate}, noting that $\rho_m =  \sum_\mu \frac{1}{d_m} \pr{\psi_{m,\mu}}$ and using the convexity of the skew-information, we thus arrive at the following relationship between the variance  in the FQH, and the variance in the partially coarse-grained FQH:
\begin{align}\label{eq:qh-inequality-1}
\var{\qc} &\leqslant \sum_m q_m \mathscr{I}_{\rho_m}(H), \nonumber \\
&  \leqslant  \sum_m q_m \sum_\mu \frac{1}{d_m} \mathscr{I}_{\pr{\psi_{m,\mu}}}(H) = \var{\qq}.
\end{align}

\subsection{Fully coarse-grained trajectories}
While the partially coarse-grained FQH has an advantage over the FQH in that it is uniquely defined by the initial state $\rho$ and the Hamiltonian $H$, it is still very operationally demanding, since the sequential measurement in \eq{eq:coarse-grained-instrument} requires full knowledge of the state $\rho$. Therefore, let us consider the case where we replace the sequential measurement by the instrument
\begin{align}\label{eq:Hamiltonian-instrument}
\ii_n(\cdot) := \Pi_n (\cdot) \Pi_n,
\end{align}
with the single measurement outcome $(n)$ defining the fully coarse-grained trajectory $\rho \mapsto \rho(n) := \Pi_n \rho \Pi_n /\tr[\Pi_n \rho]$, with probability $p(n) := \tr[\Pi_n \rho]$.    Again using the definition for conditional energy change  \eq{eq:cond-energy-change}, and the instrument \eq{eq:Hamiltonian-instrument}, we obtain the fully coarse-grained FQH to be
\begin{align}\label{eq:no-traj-quantum-heat}
\mathds{Q}(n) := \frac{\tr[H \Pi_n \rho \Pi_n]}{\tr[\Pi_n \rho]} - \frac{\tr[\Pi_n H \rho]}{\tr[\Pi_n \rho]} = 0
\end{align}
for all states $\rho$ and trajectories $(n)$, due to the fact that the projection operators $\Pi_n$ commute with the Hamiltonian $H$.  Clearly, the fully coarse-grained trajectory does not distinguish between any subspace of $\rho$, and therefore does not require any knowledge of this state; to evaluate the probabilities $p(n)$, we only need to  measure the Hamiltonian, while the initial component of the conditional energy in \eq{eq:no-traj-quantum-heat} is obtained by a weak measurement of the Hamiltonian post-selected by observing outcome $n$ of a projective measurement of the Hamiltonian,   and the second term is given by this projective energy measurement. Moreover, it is trivial that all moments of the fully coarse-grained FQH  vanish, for any state $\rho$. Consequently, we have:  $\avg{\qq} = \avg{\qc} = \avg{\mathds{Q}} = 0$ for all states $\rho$;  $\avg{\qq^k} = \avg{\qc^k} = \avg{\mathds{Q}^k} = 0$ for all $k$ if $\rho$ commutes with $H$; and 
\begin{align}
\var{\mathds{Q}} \leqslant \var{\qc} \leqslant \var{\qq}
\end{align}
for all $\rho$. 

\section{Example}

\begin{figure}[!t]
		\centering
		\includegraphics[height=0.7\linewidth]{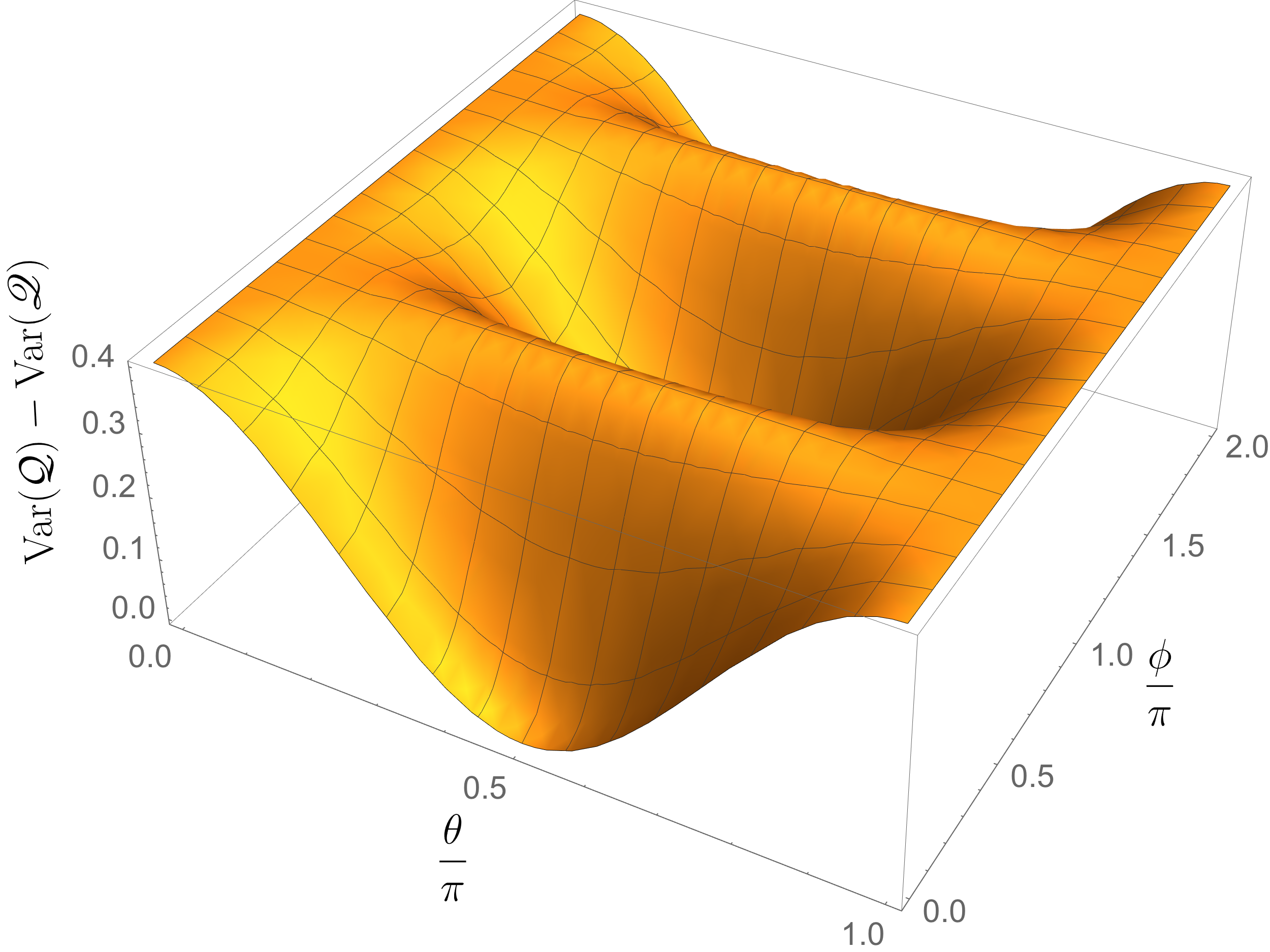}
	    \caption{The system is composed of two qubits, initially prepared in a degenerate state $\rho$ defined in \eq{eq:initial-state-example}, and then projectively measured by the  Hamiltonian defined in  \eq{eq:H-example}.  Each eigen-decomposition of  $\rho$ is parameterised by the point $(\theta, \phi) \in \re^2$, where  $\theta \in [0, \pi)$ and $\phi \in [0, 2\pi)$.  The variance in FQH,  $\var{\qq}$, depends on $(\theta, \phi)$, whereas the variance in partially coarse-grained FQH is uniquely given as $\var{\qc}=4.6$. As dictated by the inequality  \eq{eq:qh-inequality-1}, we always have $\var{\qq} \geqslant \var{\qc}$, while $\var{\qq}= \var{\qc}$  whenever $\phi = n \pi$, $n=0,1,2$, and $\theta = \pi/2$.   }
		\label{fig:bound}
	\end{figure}
Let us now illustrate our findings with a concrete example. Consider a quantum system that is composed of two qubits, with Hilbert space $\h \simeq \co^{2}\otimes \co^2$. We shall denote by $\{\ket{\pm}\}$ an orthonormal basis of $\co^2$, and 
\begin{align}
\ket{\psi_{\theta, \phi}^\pm}:= \pm e^{\pm i \phi}\cos\left(\frac{\theta}{2}\right)\ket{\pm} + \sin\left(\frac{\theta}{2}\right)\ket{\mp}
\end{align}
another orthonormal basis, with $\theta \in [0, \pi)$, $\phi \in [0, 2\pi)$ the Bloch sphere angles. Let the system have the Hamiltonian 
 \begin{align}\label{eq:H-example}
 H =  \pr{0,0} +3 \pr{0,1} + 5 \pr{1,0} + 7 \pr{1,1},
 \end{align}
where we set $\hbar =1$, with eigenstates $\{\ket{i,j} \equiv \ket{i}\otimes \ket{j}: i,j =0,1\}$, where    $\ket{1}:= \ket{\psi_{\pi/2, 0}^+}$ and $\ket{0}:= \ket{\psi_{\pi/2, 0}^-}$. Finally, let the system initially be prepared in the state 
\begin{align}\label{eq:initial-state-example}
\rho &=  \pr{+}\otimes \left(\frac{2}{10} \one \right) + \pr{-}\otimes \left(\frac{1}{10} \pr{+}  +   \frac{5}{10} \pr{-} \right), 
\end{align}
which does not commute with $H$, and has a single two-dimensional  degenerate subspace, namely the subspace projected onto by  $ \pr{+}\otimes \one$. As such, $\rho$ has infinitely many eigen-decompositions $\{\ket{+}\otimes\ket{ \psi_{\theta, \phi}^\pm}, \ket{-}\otimes\ket{\pm}\}$, which can therefore be parameterised by the Bloch sphere angles $(\theta, \phi)$. However, $\rho$ has a unique set of spectral projections, $P_1 :=  \pr{+}\otimes \one, P_2 := \pr{-,+}, P_3:= \pr{-,-}$. 

 By projectively measuring  $H$, we transform the system to $\ddh(\rho)$. The variance in the FQH $\var{\qq}$ and the partially coarse-grained FQH $\var{\qc}$ can be calculated by Eqs.  \eq{eq:variance-eigenstate} and \eq{eq:variance-coarse-grained}, respectively, and the difference between the two is plotted in \fig{fig:bound}. Clearly, the variance in the fully coarse-grained FQH is always zero.   While $\var{\qc} = 4.6$,   $\var{\qq}$ will vary depending on the degeneracy parameter $(\theta, \phi)$. However, as  dictated by  \eq{eq:qh-inequality-1}, we always have $\var{\qq} \geqslant \var{\qc}$.

\section{ Conclusions}
We examined the fluctuating quantum heat (FQH) $\qq$, as a system undergoes projective measurement with respect to its Hamiltonian $H$, within the framework of two-point eigenstate trajectories, and showed that if the initial state $\rho$ does not commute with $H$ and is degenerate, then the higher moments $\avg{\qq^k}$ will not be uniquely defined. Moreover, by showing that the FQH is an instance of increase in energy conditional on the outcome of a sequential measurement, with the first of these measurements being with respect to the eigen-decomposition of $\rho$, we provided two alternative definitions of the FQH by coarse-graining these measurements; in the partially coarse-grained FQH $\qc$, we only measure with respect to the degenerate subspaces of $\rho$, and in the fully coarse-grained FQH $\mathds{Q}$, we do not measure anything other than the Hamiltonian. As such, $\qc$ can be seen as a generalisation of $\qq$, with the two being identical when $\rho$ is non-degenerate, but the latter always providing unique higher moments, and thus being more physically meaningful. On the other hand, both $\qq$ and $\qc$ are operationally demanding, and their definition rests on full prior knowledge of the state $\rho$. $\mathds{Q}$, on the other hand, has the advantage of not requiring such prior knowledge, and is an operationally simpler alternative.

\begin{acknowledgments} The author wishes to thank M. Rafiee for insightful discussions, and acknowledges support from  the Slovak Academy of Sciences   under MoRePro project OPEQ (19MRP0027), as well as projects OPTIQUTE (APVV-18-0518) and HOQIT (VEGA 2/0161/19).
\end{acknowledgments}


\bibliographystyle{apsrev4-2}
\bibliography{CG-QH-Bib}

\begin{thebibliography}{25}%
\makeatletter
\providecommand \@ifxundefined [1]{%
 \@ifx{#1\undefined}
}%
\providecommand \@ifnum [1]{%
 \ifnum #1\expandafter \@firstoftwo
 \else \expandafter \@secondoftwo
 \fi
}%
\providecommand \@ifx [1]{%
 \ifx #1\expandafter \@firstoftwo
 \else \expandafter \@secondoftwo
 \fi
}%
\providecommand \natexlab [1]{#1}%
\providecommand \enquote  [1]{``#1''}%
\providecommand \bibnamefont  [1]{#1}%
\providecommand \bibfnamefont [1]{#1}%
\providecommand \citenamefont [1]{#1}%
\providecommand \href@noop [0]{\@secondoftwo}%
\providecommand \href [0]{\begingroup \@sanitize@url \@href}%
\providecommand \@href[1]{\@@startlink{#1}\@@href}%
\providecommand \@@href[1]{\endgroup#1\@@endlink}%
\providecommand \@sanitize@url [0]{\catcode `\\12\catcode `\$12\catcode
  `\&12\catcode `\#12\catcode `\^12\catcode `\_12\catcode `\%12\relax}%
\providecommand \@@startlink[1]{}%
\providecommand \@@endlink[0]{}%
\providecommand \url  [0]{\begingroup\@sanitize@url \@url }%
\providecommand \@url [1]{\endgroup\@href {#1}{\urlprefix }}%
\providecommand \urlprefix  [0]{URL }%
\providecommand \Eprint [0]{\href }%
\providecommand \doibase [0]{https://doi.org/}%
\providecommand \selectlanguage [0]{\@gobble}%
\providecommand \bibinfo  [0]{\@secondoftwo}%
\providecommand \bibfield  [0]{\@secondoftwo}%
\providecommand \translation [1]{[#1]}%
\providecommand \BibitemOpen [0]{}%
\providecommand \bibitemStop [0]{}%
\providecommand \bibitemNoStop [0]{.\EOS\space}%
\providecommand \EOS [0]{\spacefactor3000\relax}%
\providecommand \BibitemShut  [1]{\csname bibitem#1\endcsname}%
\let\auto@bib@innerbib\@empty
\bibitem [{\citenamefont {Seifert}(2008)}]{Seifert2008}%
  \BibitemOpen
  \bibfield  {author} {\bibinfo {author} {\bibfnamefont {U.}~\bibnamefont
  {Seifert}},\ }\href {https://doi.org/10.1140/epjb/e2008-00001-9} {\bibfield
  {journal} {\bibinfo  {journal} {Eur. Phys. J. B}\ }\textbf {\bibinfo {volume}
  {64}},\ \bibinfo {pages} {423} (\bibinfo {year} {2008})}\BibitemShut
  {NoStop}%
\bibitem [{\citenamefont {Sekimoto}(2010)}]{Sekimoto2010}%
  \BibitemOpen
  \bibfield  {author} {\bibinfo {author} {\bibfnamefont {K.}~\bibnamefont
  {Sekimoto}},\ }\href {https://doi.org/10.1007/978-3-642-05411-2} {\emph
  {\bibinfo {title} {{Stochastic Energetics}}}},\ \bibinfo {series} {Lecture
  Notes in Physics}, Vol.\ \bibinfo {volume} {799}\ (\bibinfo  {publisher}
  {Springer Berlin Heidelberg},\ \bibinfo {address} {Berlin, Heidelberg},\
  \bibinfo {year} {2010})\BibitemShut {NoStop}%
\bibitem [{\citenamefont {Seifert}(2012)}]{Seifert2012}%
  \BibitemOpen
  \bibfield  {author} {\bibinfo {author} {\bibfnamefont {U.}~\bibnamefont
  {Seifert}},\ }\href {https://doi.org/10.1088/0034-4885/75/12/126001}
  {\bibfield  {journal} {\bibinfo  {journal} {Reports Prog. Phys.}\ }\textbf
  {\bibinfo {volume} {75}},\ \bibinfo {pages} {126001} (\bibinfo {year}
  {2012})}\BibitemShut {NoStop}%
\bibitem [{\citenamefont {Campisi}\ \emph {et~al.}(2011)\citenamefont
  {Campisi}, \citenamefont {H{\"{a}}nggi},\ and\ \citenamefont
  {Talkner}}]{Campisi2011}%
  \BibitemOpen
  \bibfield  {author} {\bibinfo {author} {\bibfnamefont {M.}~\bibnamefont
  {Campisi}}, \bibinfo {author} {\bibfnamefont {P.}~\bibnamefont
  {H{\"{a}}nggi}},\ and\ \bibinfo {author} {\bibfnamefont {P.}~\bibnamefont
  {Talkner}},\ }\href {https://doi.org/10.1103/RevModPhys.83.771} {\bibfield
  {journal} {\bibinfo  {journal} {Reviews of Modern Physics}\ }\textbf
  {\bibinfo {volume} {83}},\ \bibinfo {pages} {771} (\bibinfo {year}
  {2011})}\BibitemShut {NoStop}%
\bibitem [{\citenamefont {Horowitz}\ and\ \citenamefont
  {Parrondo}(2013)}]{Horowitz2013a}%
  \BibitemOpen
  \bibfield  {author} {\bibinfo {author} {\bibfnamefont {J.~M.}\ \bibnamefont
  {Horowitz}}\ and\ \bibinfo {author} {\bibfnamefont {J.~M.~R.}\ \bibnamefont
  {Parrondo}},\ }\href {https://doi.org/10.1088/1367-2630/15/8/085028}
  {\bibfield  {journal} {\bibinfo  {journal} {New J. Phys.}\ }\textbf {\bibinfo
  {volume} {15}},\ \bibinfo {pages} {085028} (\bibinfo {year}
  {2013})}\BibitemShut {NoStop}%
\bibitem [{\citenamefont {Elouard}\ \emph
  {et~al.}(2017{\natexlab{a}})\citenamefont {Elouard}, \citenamefont
  {Bernardes}, \citenamefont {Carvalho}, \citenamefont {Santos},\ and\
  \citenamefont {Auff{\`{e}}ves}}]{Elouard2017b}%
  \BibitemOpen
  \bibfield  {author} {\bibinfo {author} {\bibfnamefont {C.}~\bibnamefont
  {Elouard}}, \bibinfo {author} {\bibfnamefont {N.~K.}\ \bibnamefont
  {Bernardes}}, \bibinfo {author} {\bibfnamefont {A.~R.~R.}\ \bibnamefont
  {Carvalho}}, \bibinfo {author} {\bibfnamefont {M.~F.}\ \bibnamefont
  {Santos}},\ and\ \bibinfo {author} {\bibfnamefont {A.}~\bibnamefont
  {Auff{\`{e}}ves}},\ }\href {https://doi.org/10.1088/1367-2630/aa7fa2}
  {\bibfield  {journal} {\bibinfo  {journal} {New J. Phys.}\ }\textbf {\bibinfo
  {volume} {19}},\ \bibinfo {pages} {103011} (\bibinfo {year}
  {2017}{\natexlab{a}})}\BibitemShut {NoStop}%
\bibitem [{\citenamefont {Manzano}\ \emph {et~al.}(2018)\citenamefont
  {Manzano}, \citenamefont {Horowitz},\ and\ \citenamefont
  {Parrondo}}]{Manzano2017}%
  \BibitemOpen
  \bibfield  {author} {\bibinfo {author} {\bibfnamefont {G.}~\bibnamefont
  {Manzano}}, \bibinfo {author} {\bibfnamefont {J.~M.}\ \bibnamefont
  {Horowitz}},\ and\ \bibinfo {author} {\bibfnamefont {J.~M.~R.}\ \bibnamefont
  {Parrondo}},\ }\href {https://doi.org/10.1103/PhysRevX.8.031037} {\bibfield
  {journal} {\bibinfo  {journal} {Phys. Rev. X}\ }\textbf {\bibinfo {volume}
  {8}},\ \bibinfo {pages} {031037} (\bibinfo {year} {2018})}\BibitemShut
  {NoStop}%
\bibitem [{\citenamefont {Elouard}\ and\ \citenamefont
  {Mohammady}(2018)}]{Elouard2018a}%
  \BibitemOpen
  \bibfield  {author} {\bibinfo {author} {\bibfnamefont {C.}~\bibnamefont
  {Elouard}}\ and\ \bibinfo {author} {\bibfnamefont {M.~H.}\ \bibnamefont
  {Mohammady}},\ }in\ \href {https://doi.org/10.1007/978-3-319-99046-0_15}
  {\emph {\bibinfo {booktitle} {Thermodyn. quantum regime Fundam. Asp. New
  Dir.}}},\ \bibinfo {series} {Fundamental Theories of Physics}, Vol.\ \bibinfo
  {volume} {195},\ \bibinfo {editor} {edited by\ \bibinfo {editor}
  {\bibfnamefont {F.}~\bibnamefont {Binder}}, \bibinfo {editor} {\bibfnamefont
  {L.~A.}\ \bibnamefont {Correa}}, \bibinfo {editor} {\bibfnamefont
  {C.}~\bibnamefont {Gogolin}}, \bibinfo {editor} {\bibfnamefont
  {J.}~\bibnamefont {Anders}},\ and\ \bibinfo {editor} {\bibfnamefont
  {G.}~\bibnamefont {Adesso}}}\ (\bibinfo  {publisher} {Springer International
  Publishing},\ \bibinfo {address} {Cham},\ \bibinfo {year} {2018})\ pp.\
  \bibinfo {pages} {363--393}\BibitemShut {NoStop}%
\bibitem [{\citenamefont {Elouard}\ \emph
  {et~al.}(2017{\natexlab{b}})\citenamefont {Elouard}, \citenamefont
  {Herrera-Mart{\'{i}}}, \citenamefont {Clusel},\ and\ \citenamefont
  {Auff{\`{e}}ves}}]{Elouard2017}%
  \BibitemOpen
  \bibfield  {author} {\bibinfo {author} {\bibfnamefont {C.}~\bibnamefont
  {Elouard}}, \bibinfo {author} {\bibfnamefont {D.~A.}\ \bibnamefont
  {Herrera-Mart{\'{i}}}}, \bibinfo {author} {\bibfnamefont {M.}~\bibnamefont
  {Clusel}},\ and\ \bibinfo {author} {\bibfnamefont {A.}~\bibnamefont
  {Auff{\`{e}}ves}},\ }\href {https://doi.org/10.1038/s41534-017-0008-4}
  {\bibfield  {journal} {\bibinfo  {journal} {npj Quantum Inf.}\ }\textbf
  {\bibinfo {volume} {3}},\ \bibinfo {pages} {9} (\bibinfo {year}
  {2017}{\natexlab{b}})}\BibitemShut {NoStop}%
\bibitem [{\citenamefont {Hayashi}\ and\ \citenamefont
  {Tajima}(2017)}]{Hayashi2017}%
  \BibitemOpen
  \bibfield  {author} {\bibinfo {author} {\bibfnamefont {M.}~\bibnamefont
  {Hayashi}}\ and\ \bibinfo {author} {\bibfnamefont {H.}~\bibnamefont
  {Tajima}},\ }\href {https://doi.org/10.1103/PhysRevA.95.032132} {\bibfield
  {journal} {\bibinfo  {journal} {Phys. Rev. A}\ }\textbf {\bibinfo {volume}
  {95}},\ \bibinfo {pages} {032132} (\bibinfo {year} {2017})}\BibitemShut
  {NoStop}%
\bibitem [{\citenamefont {Elouard}\ \emph
  {et~al.}(2017{\natexlab{c}})\citenamefont {Elouard}, \citenamefont
  {Herrera-Mart{\'{i}}}, \citenamefont {Huard},\ and\ \citenamefont
  {Auff{\`{e}}ves}}]{Elouard2017a}%
  \BibitemOpen
  \bibfield  {author} {\bibinfo {author} {\bibfnamefont {C.}~\bibnamefont
  {Elouard}}, \bibinfo {author} {\bibfnamefont {D.}~\bibnamefont
  {Herrera-Mart{\'{i}}}}, \bibinfo {author} {\bibfnamefont {B.}~\bibnamefont
  {Huard}},\ and\ \bibinfo {author} {\bibfnamefont {A.}~\bibnamefont
  {Auff{\`{e}}ves}},\ }\href {https://doi.org/10.1103/PhysRevLett.118.260603}
  {\bibfield  {journal} {\bibinfo  {journal} {Phys. Rev. Lett.}\ }\textbf
  {\bibinfo {volume} {118}},\ \bibinfo {pages} {260603} (\bibinfo {year}
  {2017}{\natexlab{c}})}\BibitemShut {NoStop}%
\bibitem [{\citenamefont {Mohammady}\ and\ \citenamefont
  {Anders}(2017)}]{Mohammady2017}%
  \BibitemOpen
  \bibfield  {author} {\bibinfo {author} {\bibfnamefont {M.~H.}\ \bibnamefont
  {Mohammady}}\ and\ \bibinfo {author} {\bibfnamefont {J.}~\bibnamefont
  {Anders}},\ }\href {https://doi.org/10.1088/1367-2630/aa8ba1} {\bibfield
  {journal} {\bibinfo  {journal} {New J. Phys.}\ }\textbf {\bibinfo {volume}
  {19}},\ \bibinfo {pages} {113026} (\bibinfo {year} {2017})}\BibitemShut
  {NoStop}%
\bibitem [{\citenamefont {Elouard}\ and\ \citenamefont
  {Jordan}(2018)}]{Elouard2018b}%
  \BibitemOpen
  \bibfield  {author} {\bibinfo {author} {\bibfnamefont {C.}~\bibnamefont
  {Elouard}}\ and\ \bibinfo {author} {\bibfnamefont {A.~N.}\ \bibnamefont
  {Jordan}},\ }\href {https://doi.org/10.1103/PhysRevLett.120.260601}
  {\bibfield  {journal} {\bibinfo  {journal} {Phys. Rev. Lett.}\ }\textbf
  {\bibinfo {volume} {120}},\ \bibinfo {pages} {260601} (\bibinfo {year}
  {2018})}\BibitemShut {NoStop}%
\bibitem [{\citenamefont {Buffoni}\ \emph {et~al.}(2019)\citenamefont
  {Buffoni}, \citenamefont {Solfanelli}, \citenamefont {Verrucchi},
  \citenamefont {Cuccoli},\ and\ \citenamefont {Campisi}}]{Buffoni2018}%
  \BibitemOpen
  \bibfield  {author} {\bibinfo {author} {\bibfnamefont {L.}~\bibnamefont
  {Buffoni}}, \bibinfo {author} {\bibfnamefont {A.}~\bibnamefont {Solfanelli}},
  \bibinfo {author} {\bibfnamefont {P.}~\bibnamefont {Verrucchi}}, \bibinfo
  {author} {\bibfnamefont {A.}~\bibnamefont {Cuccoli}},\ and\ \bibinfo {author}
  {\bibfnamefont {M.}~\bibnamefont {Campisi}},\ }\href
  {https://doi.org/10.1103/PhysRevLett.122.070603} {\bibfield  {journal}
  {\bibinfo  {journal} {Phys. Rev. Lett.}\ }\textbf {\bibinfo {volume} {122}},\
  \bibinfo {pages} {070603} (\bibinfo {year} {2019})}\BibitemShut {NoStop}%
\bibitem [{\citenamefont {Solfanelli}\ \emph {et~al.}(2019)\citenamefont
  {Solfanelli}, \citenamefont {Buffoni}, \citenamefont {Cuccoli},\ and\
  \citenamefont {Campisi}}]{Solfanelli2019}%
  \BibitemOpen
  \bibfield  {author} {\bibinfo {author} {\bibfnamefont {A.}~\bibnamefont
  {Solfanelli}}, \bibinfo {author} {\bibfnamefont {L.}~\bibnamefont {Buffoni}},
  \bibinfo {author} {\bibfnamefont {A.}~\bibnamefont {Cuccoli}},\ and\ \bibinfo
  {author} {\bibfnamefont {M.}~\bibnamefont {Campisi}},\ }\href
  {https://doi.org/10.1088/1742-5468/ab3721} {\bibfield  {journal} {\bibinfo
  {journal} {J. Stat. Mech. Theory Exp.}\ }\textbf {\bibinfo {volume} {2019}},\
  \bibinfo {pages} {0} (\bibinfo {year} {2019})}\BibitemShut {NoStop}%
\bibitem [{\citenamefont {Santos}\ \emph {et~al.}(2019)\citenamefont {Santos},
  \citenamefont {C{\'{e}}leri}, \citenamefont {Landi},\ and\ \citenamefont
  {Paternostro}}]{Santos2019}%
  \BibitemOpen
  \bibfield  {author} {\bibinfo {author} {\bibfnamefont {J.~P.}\ \bibnamefont
  {Santos}}, \bibinfo {author} {\bibfnamefont {L.~C.}\ \bibnamefont
  {C{\'{e}}leri}}, \bibinfo {author} {\bibfnamefont {G.~T.}\ \bibnamefont
  {Landi}},\ and\ \bibinfo {author} {\bibfnamefont {M.}~\bibnamefont
  {Paternostro}},\ }\href {https://doi.org/10.1038/s41534-019-0138-y}
  {\bibfield  {journal} {\bibinfo  {journal} {npj Quantum Inf.}\ }\textbf
  {\bibinfo {volume} {5}},\ \bibinfo {pages} {23} (\bibinfo {year}
  {2019})}\BibitemShut {NoStop}%
\bibitem [{\citenamefont {Mohammady}\ \emph {et~al.}(2020)\citenamefont
  {Mohammady}, \citenamefont {Auff{\`{e}}ves},\ and\ \citenamefont
  {Anders}}]{Mohammady2019d}%
  \BibitemOpen
  \bibfield  {author} {\bibinfo {author} {\bibfnamefont {M.~H.}\ \bibnamefont
  {Mohammady}}, \bibinfo {author} {\bibfnamefont {A.}~\bibnamefont
  {Auff{\`{e}}ves}},\ and\ \bibinfo {author} {\bibfnamefont {J.}~\bibnamefont
  {Anders}},\ }\href {https://doi.org/10.1038/s42005-020-0356-9} {\bibfield
  {journal} {\bibinfo  {journal} {Commun. Phys.}\ }\textbf {\bibinfo {volume}
  {3}},\ \bibinfo {pages} {89} (\bibinfo {year} {2020})}\BibitemShut {NoStop}%
\bibitem [{\citenamefont {Mohammady}\ and\ \citenamefont
  {Romito}(2019)}]{Mohammady2019c}%
  \BibitemOpen
  \bibfield  {author} {\bibinfo {author} {\bibfnamefont {M.~H.}\ \bibnamefont
  {Mohammady}}\ and\ \bibinfo {author} {\bibfnamefont {A.}~\bibnamefont
  {Romito}},\ }\href {https://doi.org/10.22331/q-2019-08-19-175} {\bibfield
  {journal} {\bibinfo  {journal} {Quantum}\ }\textbf {\bibinfo {volume} {3}},\
  \bibinfo {pages} {175} (\bibinfo {year} {2019})}\BibitemShut {NoStop}%
\bibitem [{\citenamefont {Heinosaari}\ and\ \citenamefont
  {Ziman}(2011)}]{Heinosaari2011}%
  \BibitemOpen
  \bibfield  {author} {\bibinfo {author} {\bibfnamefont {T.}~\bibnamefont
  {Heinosaari}}\ and\ \bibinfo {author} {\bibfnamefont {M.}~\bibnamefont
  {Ziman}},\ }\href {https://doi.org/10.1017/CBO9781139031103} {\emph {\bibinfo
  {title} {{The Mathematical language of Quantum Theory}}}}\ (\bibinfo
  {publisher} {Cambridge University Press},\ \bibinfo {address} {Cambridge},\
  \bibinfo {year} {2011})\BibitemShut {NoStop}%
\bibitem [{\citenamefont {Busch}\ \emph {et~al.}(2016)\citenamefont {Busch},
  \citenamefont {Lahti}, \citenamefont {Pellonp{\"{a}}{\"{a}}},\ and\
  \citenamefont {Ylinen}}]{Busch2016a}%
  \BibitemOpen
  \bibfield  {author} {\bibinfo {author} {\bibfnamefont {P.}~\bibnamefont
  {Busch}}, \bibinfo {author} {\bibfnamefont {P.}~\bibnamefont {Lahti}},
  \bibinfo {author} {\bibfnamefont {J.-P.}\ \bibnamefont
  {Pellonp{\"{a}}{\"{a}}}},\ and\ \bibinfo {author} {\bibfnamefont
  {K.}~\bibnamefont {Ylinen}},\ }\href
  {https://doi.org/10.1007/978-3-319-43389-9} {\emph {\bibinfo {title}
  {{Quantum Measurement}}}},\ Theoretical and Mathematical Physics\ (\bibinfo
  {publisher} {Springer International Publishing},\ \bibinfo {address} {Cham},\
  \bibinfo {year} {2016})\BibitemShut {NoStop}%
\bibitem [{\citenamefont {Aharonov}\ \emph {et~al.}(1988)\citenamefont
  {Aharonov}, \citenamefont {Albert},\ and\ \citenamefont
  {Vaidman}}]{Aharonov1988}%
  \BibitemOpen
  \bibfield  {author} {\bibinfo {author} {\bibfnamefont {Y.}~\bibnamefont
  {Aharonov}}, \bibinfo {author} {\bibfnamefont {D.~Z.}\ \bibnamefont
  {Albert}},\ and\ \bibinfo {author} {\bibfnamefont {L.}~\bibnamefont
  {Vaidman}},\ }\href {https://doi.org/10.1103/PhysRevLett.60.1351} {\bibfield
  {journal} {\bibinfo  {journal} {Phys. Rev. Lett.}\ }\textbf {\bibinfo
  {volume} {60}},\ \bibinfo {pages} {1351} (\bibinfo {year}
  {1988})}\BibitemShut {NoStop}%
\bibitem [{\citenamefont {Haapasalo}\ \emph {et~al.}(2011)\citenamefont
  {Haapasalo}, \citenamefont {Lahti},\ and\ \citenamefont
  {Schultz}}]{Haapasalo2011}%
  \BibitemOpen
  \bibfield  {author} {\bibinfo {author} {\bibfnamefont {E.}~\bibnamefont
  {Haapasalo}}, \bibinfo {author} {\bibfnamefont {P.}~\bibnamefont {Lahti}},\
  and\ \bibinfo {author} {\bibfnamefont {J.}~\bibnamefont {Schultz}},\ }\href
  {https://doi.org/10.1103/PhysRevA.84.052107} {\bibfield  {journal} {\bibinfo
  {journal} {Phys. Rev. A}\ }\textbf {\bibinfo {volume} {84}},\ \bibinfo
  {pages} {052107} (\bibinfo {year} {2011})}\BibitemShut {NoStop}%
\bibitem [{\citenamefont {Wigner}\ and\ \citenamefont
  {Yanase}(1963)}]{Wigner1963}%
  \BibitemOpen
  \bibfield  {author} {\bibinfo {author} {\bibfnamefont {E.~P.}\ \bibnamefont
  {Wigner}}\ and\ \bibinfo {author} {\bibfnamefont {M.~M.}\ \bibnamefont
  {Yanase}},\ }\href {https://doi.org/10.1073/pnas.49.6.910} {\bibfield
  {journal} {\bibinfo  {journal} {Proc. Natl. Acad. Sci.}\ }\textbf {\bibinfo
  {volume} {49}},\ \bibinfo {pages} {910} (\bibinfo {year} {1963})}\BibitemShut
  {NoStop}%
\bibitem [{\citenamefont {Lieb}(1973)}]{Lieb1973}%
  \BibitemOpen
  \bibfield  {author} {\bibinfo {author} {\bibfnamefont {E.~H.}\ \bibnamefont
  {Lieb}},\ }\href {https://doi.org/10.1016/0001-8708(73)90011-X} {\bibfield
  {journal} {\bibinfo  {journal} {Adv. Math. (N. Y).}\ }\textbf {\bibinfo
  {volume} {11}},\ \bibinfo {pages} {267} (\bibinfo {year} {1973})}\BibitemShut
  {NoStop}%
\bibitem [{\citenamefont {Takagi}(2019)}]{Takagi2018}%
  \BibitemOpen
  \bibfield  {author} {\bibinfo {author} {\bibfnamefont {R.}~\bibnamefont
  {Takagi}},\ }\href {https://doi.org/10.1038/s41598-019-50279-w} {\bibfield
  {journal} {\bibinfo  {journal} {Sci. Rep.}\ }\textbf {\bibinfo {volume}
  {9}},\ \bibinfo {pages} {14562} (\bibinfo {year} {2019})}\BibitemShut
  {NoStop}%
\end{thebibliography}%

\end{document}